\documentstyle[12pt]{article}
\def\a{\begin{eqnarray}}
\def\b{\end{eqnarray}}
\def\0{\nonumber}
\def\ba{\begin{array}}
\def\ea{\end{array}}

\def\q{{\bar{\cal Q}}}

\def\al{{\alpha}}
\def\lm{{\lambda}}

\renewcommand{\theequation}{\thesection.\arabic{equation}}

\newlength{\extraspace}
\newlength{\extraspaces}
\newcounter{dummy}
\newcommand{\ai}{
\addtocounter{equation}{1}
\setcounter{dummy}{\value{equation}}
\setcounter{equation}{0}
\renewcommand{\theequation}{\thesection.\arabic{dummy}\alph{equation}}
\begin{eqnarray}
\addtolength{\abovedisplayskip}{\extraspaces}
\addtolength{\belowdisplayskip}{\extraspaces}
\addtolength{\abovedisplayshortskip}{\extraspace}
\addtolength{\belowdisplayshortskip}{\extraspace}}
\newcommand{\bj}{
\end{eqnarray}
\setcounter{equation}{\value{dummy}}
\renewcommand{\theequation}{\thesection.\arabic{equation}}}
\def\d{{\partial}}

\begin{document}
\begin{flushright}
hep-th/yymmxxx
\end{flushright}
\vskip0.5cm
\centerline{\LARGE\bf Quantum Chaos in Multi--Matrix Models }
\vskip0.3cm
\centerline{\large   E.Vinteler}
\centerline{International School for Advanced Studies (SISSA/ISAS)}
\centerline{Via Beirut 2, 34014 Trieste, Italy}
\vskip0.5cm
\vskip5cm
\abstract{We propose a possible resolution for  the  problem of why  the
semicircular law is not observed, whilst the random matrix  hypothesis
describes well the fluctuation of energy spectra. We show in the  random
2-matrix model that the interactions between  the  quantum  subsystems
alter the semicircular law of level density. We  consider  also  other
types of interactions in the chain- and star-multimatrix  models.  The
connection with the Calogero-Sutherland models is briefly discussed.}
\vfill\eject

\section{Introduction}
In heavy nuclei,the complicated many-body interactions lead to statistical
theories which explain only the average properties.
One of these theories is the random matrix hypothesis
\cite{Porter}\cite{Mehta1}.
It supposes that the
nuclear hamiltonian in a arbitrary basis of functions is a $N\times N$
matrix with $N$ large and elements distributed at random.
The joint probability function of the eigenvalues $\lm_1,\ldots \lm_N$
 of this  matrix model is given by:
\a
P(\lm_1,\ldots \lm_N)=\exp(-\sum_{i=1}^N\lm_i^2)\prod_{i<j}(\lm_i-\lm_j)^\beta
\label{1mat}
\b
where $\beta=1,2,4$ for orthogonal, hermitean and,respectively unitary
ensembles.
Integrating over eigenvalues $\lm_{k+1}\ldots \lm_N$ we get the joint
distribution function for few levels:
\a
P(\lm_1,\ldots \lm_k)=\int d\lm_{k+1}\ldots d\lm_N P(\lm_1,\ldots \lm_N)
\b
All these joint distribution functions can be expressed in terms of the
Dyson correlation function $K(\lm_1,\lm_2)$:
\a
P(\lm_1,\ldots \lm_k)=\sum_\sigma (-1)^\sigma K(\lm_1,\lm_{\sigma_1})
\ldots K(\lm_k,\lm_{\sigma_k})\0
\b
where $\sigma$ is the permutation of k levels.In the special case $k=1$ the
Dyson correlation function coincides with level density $K(\lm,\lm)=P(\lm)$.

The density of levels for the 1-matrix model satisfies the semicircular law:
\a
P(\lm)= \sqrt{\beta N/2- \lm^2}\0
\b
and the Dyson correlation function behaves as $(\sigma \ll \lm)$:
\a
K(\lm-{1\over 2}\sigma,\lm+{1\over 2}\sigma) \simeq {\sin(\pi \sigma
P(\lm))\over \pi \sigma
(\beta N/2)}\0
\b

The Dyson correlation function describes  well  the  fluctuations  of
quantum  systems,but  the  semicircular  law  is  not   observed   in   the
experimental data for the density of levels. A  possible  resolution  of
problem is to consider instead one random matrix few random matrices  in
interaction. As  we  will  see  ,  even  a  small  interaction  gives  a
calitatively new behaviour for the level density.

As an interesting generalization of the random matrix hypothesis is to
consider $q$
matrices describing $q$ nuclear systems in interaction.The total action of
such system is:
\a
S_1=\sum_{\al=1}^q\sum_{i=1}^N(t_\al(\lm_i^{(\al)})^2+u_\al\lm_i^{(\al)}) +
\sum_{\al=1}^{q-1}\sum_{i=1}^Nc_\al\lm_i^{(\al)}\lm_i^{(\al+1)}
\b
This system describes a chain of matrices with neighbour interaction.We can
add a term describing the two-body interaction of constituent nuclear
subsystems:
$$\sum_{|\al-\beta|\neq
1}\sum_{i=1}^Nc_{\al,\beta}\lm_i^{(\al)}\lm_i^{(\beta)}$$
We have different sets of energy levels $\lm_1^{(\al)},\ldots \lm_N^{(\al)},
\al=1\ldots q$ with distribution probability:
\a
P(\lm_1^{(1)},\ldots \lm_N^{(1)}\ldots \lm_1^{(q)},\ldots \lm_N^{(q)})=
\exp(S)\prod_{i<j}(\lm_i^{(1)}-\lm_j^{(1)})(\lm_i^{(q)}-\lm_j^{(q)})
\b
We have level repulsion only for the first and last energy level set.Hence for
this model the intermediate energy level sets are "classical" and interact
with "quantum" first and last energy level sets.Integrating over all
intermediate matrices we remain with a two-matrix model.

Kharchev and others have
considered the so-called conformal matrix models that contain additional
repulsion terms also for intermediate matrices \cite{Kharchev}.

Another special random matrix model is the star-matrix model having the action:
\a
S_2=\sum_{i=1}^N(t_0(\lm_i^{(0)})^2+u_0\lm_i^{(0)})
+\sum_{\al=1}^q\sum_{i=1}^N(t_\al(\lm_i^{(\al)})^2+u_\al\lm_i^{(\al)}) +
\sum_{\al=1}^{q}\sum_{i=1}^Nc_\al\lm_i^{(\al)}\lm_i^{(0)}
\b
The joint distribution of this model reduces  again to that of 2-matrix model.

\section{Quantum Chaos in two-matrix model}
\setcounter{equation}{0}
\setcounter{subsection}{0}

We introduce the distribution probability:
\a
P(\lm_1\ldots \lm_N,\mu_1\ldots \mu_N)=\exp\sum_{i=1}^N(V_1(\lm_i)+
V_2(\mu_i)+c_i\lm_i\mu_i)\prod_{i<j}(\lm_i-\lm_j)(\mu_i-\mu_j)
\label{2mat}
\b
with $V_\al(\tau)=t_\al\tau^2+u_\al\tau,\al=1,2 $ and joint distribution
function :
\a
P(\lm_1\ldots \lm_i,\mu_1\ldots \mu_j)=\int d\lm_{i+1}\ldots d\lm_N,
d\mu_{j+1}\ldots d\mu_N P(\lm_1\ldots \lm_N,\mu_1\ldots \mu_N)
\b

We show that the level densities $P(\lm),P(\mu)$ and the
joint probability distributions
$P(\lm_1,\lm_2),P(\mu_1,\mu_2)$ are exactly like those of the hermitean
1-matrix model with distribution probability (\ref{1mat}):
\a
P(\lm)=P_{Herm}(\lm'),P(\mu)=P_{Herm}(\mu')\\
P(\lm_1,\lm_2)\sim P_{Herm}(\lm_1',\lm_2'),P(\mu_1,\mu_2)\sim P_{Herm}
(\mu_1',\mu_2')\0
\b
The new joint probability distributions $P(\lm,\mu)$ behaves in a different way
because
 we have not  energy repulsion between levels of different sets.

If we set from beginning the coupling $c=0$
we get two independent orthogonal 1-matrix models and we have:
$$P(\lm,\mu)=P_{Orth}(\lm')P_{Orth}(\mu') $$

For $c\neq 0$, $P(\lm,\mu)$ behaves like the 1-matrix  Dyson  correlation
function:
$$P(\lm,\mu)\sim K(\lm,\mu)$$
When $c  \rightarrow  0$,$P(\lm,\mu)$ does not split in  two  orthogonal
1-matrix models.

Here $\lm',\mu'$ are related with the coefficients of the $Q$-matrices.
\a
\lm'={\lm-a_0\over \sqrt{2a_1}},\mu'={\mu-b_0\over \sqrt{2b_1}}
\b
with:
\a
a_0 = {c_1u_2-2t_2 u_1 \over 4t_1t_2 -c_1^2},
b_0=   {c_1u_1-2t_1 u_2 \over 4t_1t_2 -c_1^2},\0\\
a_1= -{2t_2  \over 4t_1t_2 -c_1^2},
b_1=-{2t_1  \over 4t_1t_2 -c_1^2}
\b

In the rest of the section we demonstrate the above relations.
We introduce the  $Q$-matrices,which in the two-matrix case  have the form:
\a
Q_1=I_++a_0I_0+a_1\epsilon_-,
{\overline Q_2}=I_++b_0I_0+b_1\epsilon_-,
\b
with $$I_+=\sum_{n=0}^NE_{n,n+ 1},I_0=\sum_{n=0}^NE_{n,n},
\epsilon_-=\sum_{n=0}^NnE_{n,n- 1}$$

The $Q$ matrices are defined as:
\a
\int d\lambda d\mu \xi_n(\lambda)
\lambda_{\alpha}e^{V_1+V_2+c\lm\mu} \eta_m(\mu)=Q_{\al,nm}h_m,\al=1,2 \0
\b
where $h_n=h_0 R^n$ and $R=c/(4t_2t_1-c^2)$.

 $\xi,\eta$ are  orthogonal polynomials
\a
\xi_n(\lambda_1)=\lambda_1^n+\ldots,\
\eta_n(\mu)=\mu^n+\ldots \0
\b
 satisfying the orthogonality condition:
\a
\int d\lambda d\mu \xi_n(\lambda)e^{V_1+V_2+c\lm\mu} \eta_m(\mu)
=h_n\delta_{nm}
\b
{}From the definition of $Q$-matrices ($Q_{1,mn}\xi_m=\lm_n\xi_n,
{\overline Q_{2,mn}}\eta_m=\mu_n\eta_n$
we have the following
recursion relations of the orthogonal polynomials:
\a
\lm\xi_n(\lm)=\xi_{n+1}(\lm)+a_0\xi_n(\lm)+a_1\xi_{n-1}(\lm)\0\\
\mu\eta_n(\mu)=\eta_{n+1}(\mu)+b_0\eta_n(\mu)+b_1\eta_{n-1}(\mu)
\b
Solving these recursion relations it follows that  $\xi,\eta$ are Hermite
functions:
\a
 \xi_n(\lm)=\al_n H_n(\lm'),
\eta_m(\mu)=\beta_m H_m(\mu')\0
\b
To get the proportionality coefficients $\al_n,\beta_m$ we use the
orthogonality relation and the Gauss transform:
\a
(2\pi u)^{-1/2}\int dy e^{-(x-y)^2/(2u)}H_n(y)=(1-2u)^{n/2}
H_n((1-2u)^{-1/2}x)
\b
Writing the action as:
\a
S&=&V_1(\lm)+V_2(\mu)+c\lm\mu=t_1\lm^2+u_1\lm+t_2\mu^2+u_2\mu+c\lm\mu=\0\\
&=&S_0+t_2\left(\mu+{u_2+c\lm\over 2t_2}\right)^2-
\left({\lm-a_0\over \sqrt{2a_1}}\right)^2
\b
with:
\a
S_0=-{t_1u_2^2+t_2u_1^2-cu_1u_2\over 4t_1t_2-c^2}
\b
we have:
\a
\int d\lm d\mu \xi_n(\lambda)e^{V_1+V_2+c\lm\mu} \eta_m(\mu)&=&
\al_n\beta_m \delta_{nm}e^{S_0}{2\pi \over \sqrt{4t_1t_2-c^2}}
({c\over\sqrt{4t_1t-2}})^n 2^n n!=\0\\
&=&h_0 \delta_{nm}R^n=h_0\left({c\over 4t_1t_2-c^2}\right)^n
\b
In conclusion:
\a
 \xi_n(\lm)=(2\pi n!)^{-1/2}2^{-n/2}(\sqrt{2a_1})^n H_n(\lm'),\\
\eta_m(\mu)=(2\pi n!)^{-1/2}2^{-m/2}(\sqrt{2b_1})^n H_m(\mu')\0
\label{pol}
\b
and:
\a
h_0=(4t_1t_2-c^2)^{-1/2}\exp(S_0)\0
\b

We can now calculate the joint probability distribution $P(\lm,\mu)$.Because we
can write the
two Vandermonde determinants in terms of orthogonal polynomials $\xi_n,\eta_m$

\a
\Delta(\lm)\Delta(\mu)=\sum_n\xi_n(\lm_1)\Xi_n(\lm_2\ldots \lm_N)
\sum_m\eta_m(\mu_1)\Theta_m(\mu_2\ldots \mu_N)\0
\b
and the algebraic complements satisfy:
\a
\int \prod_{i=2}^N(d\lm_i d\mu_i)\Xi_n(\lm_2\ldots \lm_N)
\Theta_m(\mu_2\ldots \mu_N)=(N-1)!\delta_{nm}\0
\b
we get for joint probability distribution:
\a
P(\lm,\mu)={1\over N}e^{S}\sum_{n=0}^{N-1}h_n^{-1}\xi_n(\lm)\eta_n(\mu)
\label{214}
\b

It is easy to derive the expression for symmetric joint distribution of
pairs of eigenvalues in terms of $P(\lm,\mu)$:
\a
P(\lm_1,\ldots,\lm_k,\mu_1,\ldots,\mu_k)=\sum_\sigma (-1)^\sigma
P(\lm_1,\mu_{\sigma_1})\ldots P(\lm_k,\mu_{\sigma_k})\0
\b
Integrating in $\lm_{j+1}\ldots \lm_k$ we obtain the asymmetric
joint distribution of eigenvalues:
\a
P(\lm_1,\ldots,\lm_j,\mu_1,\ldots,\mu_k)=\sum_\sigma (-1)^\sigma
P(\lm_1,\mu_{\sigma_1})\ldots P(\lm_j,\mu_{\sigma_j})
P(\mu_{\sigma_{j+1}})P(\mu_{\sigma_k})\0
\b

In the limit of large $N$ we have the usual behaviour of
semi-circular law:
\a
P(\lm)=\sqrt{2N-\lm'^2},P(\mu)=\sqrt{2N-\mu'^2}\0
\b

To calculate  the joint distribution of two  eigenvalues $P(\lm,\mu)$
in the large $N$ limit we associate it with the quantum mechanical system :

\a
[{1\over
2}(p_\lm^2+p_\mu^2)+V_1(\lm)+V_2(\mu)+c\lm\mu]\phi_n(\lm)\psi_m(\mu)=E_{nm}
\phi_n(\lm)\psi_m(\mu)\0
\b
where $p_\lm=i\d/\d\lm,p_\mu=i\d/\d\mu$ are the usual momenta operators and
\a
\phi_n(\lm)=\exp(-\lm'^2/2)\eta_n(\lm)\0\\
\psi_m(\mu)=\exp(-\mu'^2/2)\xi_m(\mu)
\b

For $c=0$ we get two decoupled quantum systems:
\a
(p_\lm^2+\lm'^2)\phi_n(\lm)=2E_{1,n}\phi_n(\lm)\0\\
(p_\mu^2+\mu'^2)\psi_m(\mu)=2E_{2,m}\psi_m(\mu)
\label{2schro}
\b
where $E_{nm}=E_{1,n}+E_{2,m}$.

In the large $N$ limit $E_{nm}$ behaves like
$\sim N$ and because we are searching for symmetric solutions we have
$E_{1,n}=E_{2,m}\sim N/2$.The joint distribution of two  eigenvalues
 $P(\lm,\mu)$ will be:
\a
P(\lm,\mu)=\sqrt{2E_{1,n}-\lm'^2}\sqrt{2E_{2,n}-\mu'^2}
\label{2density}
\b
or
\a
P(\lm,\mu)=\sqrt{N-\lm'^2}\sqrt{N-\mu'^2}
\label{2density'}
\b
We can see that for $c=0$, $P(\lm,\mu)$ is the product of density energy
levels for orthogonal anssembles.If we integrate the last matrix ,we get the
1-matrix model .In our case this is equivalent with the condition
$2E_{2,m}=p_\mu^2+V_2(\mu)=0$ in (\ref{2schro}) or in other words
the second system has no
contribution in the joint distribution of two  eigenvalues.
The equation (\ref{2density}) is replaced by:
\a
P(\lm)=\sqrt{2N-\lm'^2}\0
\b

For $c\neq 0$, after summing relation (\ref{214}) and using the  asymptotic
formula ($n$ large) for the Hermite polynomial (near origin):
\a
H_n{}=e^{x^2}{\Gamma  (n+1)\over\Gamma  (n/2+1)}\cos  (\sqrt{2n+1}-{n\pi
\over 2})+O(1/\sqrt{n})\0
\b
we obtain (up the exponent $S+(\lm'^2+\mu'^2)/2$):
\a
P(\lm,\mu)\sim  {\sin\sqrt{2N}(\lm'-\mu')\over
\pi  N(\lm'-\mu')}, \quad \lm,\mu \ \ \mbox{near} \ \ 0
\b
We also get for arbitrary $\lm, \mu,\lm\ll\mu$:
\a
P(\lm,\mu)\sim  {\sin\sqrt{2N-(\al\lm)^2}\epsilon (\al\lm)\over
\pi  N \epsilon  (\al\lm)}
\b
where:
\a
\epsilon &=&{1\over \sqrt{2a_1}}-{1\over \sqrt{2b_1}}\0\\
\al &=&{1\over 2}\left({1\over \sqrt{2a_1}}-{1\over \sqrt{2b_1}}\right)
\b

For the asymmetric potential $t_1=1/(a+\tau)^2,t_2=1/(a-\tau)^2, (\tau \ll a)$
and   a   small   interaction   $c\approx  0$,   we   have   $\epsilon\sim
\tau/a^2,\al\sim 1/(2a)$ and $\epsilon \ll \al$. When $\tau  \rightarrow  0$
(symmetric potential) $P(\lm,\mu\sim \lm)$ tends to the level density of
hermitean 1-matrix model $P_{Herm}(\lm)$. The interaction (even a  small
one) of asymmetric energy levels changes dramatically  the level density
$P(\lm,\lm) $ of the system.

If for $\tau  \rightarrow  0$ we get the usual semicircular law, a small
asymmetry creates some peaks in the level  density  $P(\lm,\lm)  $  (see
figure  1).  The  observed  behaviour  is  the  quantum  analog for
chaotical behaviour  of  two interacting classical oscilators.

\section{$q$-matrix model}
\setcounter{equation}{0}
\setcounter{subsection}{0}

As a random $q$-multimatrix model we choose the one with partition function:
\a
Z=\int \prod_{\al=1}^q \lm_\al \Delta(\lm_1) \Delta(\lm_q)
\exp( \sum_{\al =1}^q t_\al \lambda_\al^2 + \sum_{\al =1}^{q-1}
c_\al \lambda_\al \lambda_{\al +1})
\b
We show that the joint probability is :
\a
P(\lm_\al,\lm_\beta)=P_{Herm}(\lm_\al',\lm_\beta'),1\leq \al\leq\beta\leq q
\label{qprob}
\b
where:
\a
\lm_\al'=\lm_\al/\sqrt{2a_\al},\0
\b
The parameters $a_\al$ are the coefficients of the $Q$-matrices.

 The $Q(\al)$ have only three non--vanishing
diagonal lines, the main diagonal and the two adjacent lines.
\a
Q(\al) =b_\al I_+ +a_\al \epsilon_{-}\label{Qalpha}
\b
where in the particular cases we know that $b_1 =1$ and $a_q = R$.
We can write the parameters in terms of the determinants of two matrices (we
use the results of the paper \cite{BNV}):
\a
&&b_\al = (-1)^\al (c_1c_2\ldots c_{\al-1})^{-1} \det X_{\al-1}\0\\
&& R = (-1)^q  c_1 c_2 \ldots c_{q-1} \Big(\det X_q\Big)^{-1}
\label{sollinsys}\\
&&a_\al = (-1)^\al  c_1c_2\ldots c_{\al-1} \frac{\det Y_{\al+1}}{\det X_q}\0
\b
The matrices $X_\al$ and $Y_\al$,are  defined as follows
\a
X_\al = \left(\ba{cccccc} 2t_1 & c_1 & 0 &\ldots& 0 & 0 \\
                         c_1 & 2t_2 & c_2 &\ldots &0&0\\
                          0 &c_2 & 2t_3 &\ldots & 0&0\\
                          \ldots&\ldots&\ldots&\ldots&\ldots&\ldots\\
                          0&0&0& \ldots& 2t_{\al-1}& c_{k-1}\\
                          0&0&0& \ldots& c_{k-1}& 2t_\al\ea\right)\label{Xk}
\b
and
\a
Y_\al = \left(\ba{cccccc} 2t_\al & c_k & 0 &\ldots& 0 & 0 \\
                         c_k & 2t_{\al+1} & c_{k+1} &\ldots &0&0\\
                          0 &c_{k+1} & 2t_{\al+2} &\ldots & 0&0\\
                          \ldots&\ldots&\ldots&\ldots&\ldots&\ldots\\
                          0&0&0& \ldots& 2t_{q-1}& c_{q-1}\\
                          0&0&0& \ldots& c_{q-1}& 2t_q\ea\right)\label{Yk}
\b
Of course $Y_1\equiv X_q$.

 As we made before for the 2-matrix model we introduce
 the orthogonal polynomials
\a
\xi_n(\lambda_1)=\lambda_1^n+\hbox{lower powers},\qquad\qquad
\eta_n(\lambda_q)=\lambda_q^n+\hbox{lower powers}\0
\b
which satisfy the orthogonality relations
\a
\int  d\lambda_1\ldots d\lambda_q\xi_n(\lambda_1)
e^{S}
\eta_m(\lambda_q)=h_n\delta_{nm}\label{orth1}
\b
where
\a
S=\sum_{\al=1}^{q}t_{\al}\lambda_{\al}^2
+\sum_{\al=1}^{q-1}c_{\al}\lambda_{\al}\lambda_{\al+1}.\0
\b
We introduce also the basic intermediate functions:
\a
\xi^{(\al)}_n(\lm_{\al})\equiv\int\prod_{\beta=1}^{\al-1}
d\lm_{\beta}\xi_n(\lambda_1)e^{S_{\al}}.\label{xial}
\b
and
\a
\eta^{(\al)}_n(\lm_{\al})\equiv\int\prod_{\beta=\al+1}^{q}
d\lm_{\beta}e^{S_{\al}'}
\eta_m(\lambda_q).\label{etaal}
\b
where we denote
\a
&&S_{\al}=\sum_{\beta=1}^{\al-1}t_{\beta}
\lambda_{\beta}^2
+\sum_{\beta=1}^{\al-1}c_{\beta}\lambda_{\beta}\lambda_{\beta+1}.\0\\
&&S_{\al}'=\sum_{\beta=\al+1}^{q-1}t_{\beta}
\lambda_{\beta}^2
+\sum_{\beta=\al}^{q-1}c_{\beta}\lambda_{\beta}\lambda_{\beta+1}.\0
\b
Obviously we have
\a
\xi^{(1)}_n(\lm_{1})=\xi_n(\lambda_1),\qquad\qquad
\eta^{(q)}_n(\lm_{q})=\eta_m(\lambda_q).\0
\b
In the general case when we have arbitrary potentials
 one sees immediately that $\xi^{(\al)}$
and $\eta^{(\al)}$ are not polynomials anymore.In our case of gaussian
potentials
these intermediate functions are again Hermite functions ,but with different
arguments.
However they still satisfy an orthogonality relation
\a
\int \prod_{\gamma=\al}^\beta d\lm_{\gamma}\xi^{(\al)}_n(\lm_{\al})
e^{S-S_\al-S_\al'}(\lm_{\al})\eta^{(\beta)}_m(\lm_{\beta})
=\delta_{nm}h_n,\qquad  1\leq\al\leq\beta\leq q.\label{orth3}
\b
These basic intermediate functions permit to write the intermediate $Q$
matrices as:
\a
\int d\lm_{\al}\xi^{(\al)}_n(\lm_{\al})
e^{V_{\al}(\lm_{\al})}\lm_{\al}\eta^{(\al)}_n(\lm_{\al})
=Q_{\al,nm}h_m= \q_{\al,nm}h_n,\qquad  1\leq\al\leq q.\label{qal'}
\b
The equations satisfied by  basic intermediate functions are:
\a
&&\lm_{\al}\xi^{(\al)}=Q_\al\xi^{(\al)},\qquad 1\leq\al\leq q.\label{specal}\\
&&\lm_{\al}\eta^{(\al)}=\q_\al\eta^{(\al)},\qquad 1\leq\al\leq q.
\label{specal'}
\b
These equations together with the explicit form of $Q$-matrices permits to find
the
basic intermediate functions $\xi_n^{(\al)},\ \eta_m^{(\al)}$:
\a
\lm_\al\xi_n^{(\al)}(\lm_\al)&=&b_\al\xi_{n+1}(\lm_\al)+a_\al\xi_{n-1}(\lm_\al)\0\\
\lm_\al\eta_n^{(\al)}(\lm_\al)&=&(a_\al/R) \eta_{n+1}(\lm_\al)+b_\al
R\eta_{n-1}(\lm_\al)
\b
Solving these recursion relations it follows that  $\xi^{(\al)},\eta^{(\al)}$
 are Hermite functions for gaussian potentials:
\a
 \xi_n^{(\al)}(\lm_\al)= (2\pi n!)^{-1/2}2^{-n/2}(\sqrt{2a_\al})^n
H_n(\lm_\al'),\0\\
\eta_m^{(\al)}(\lm_\al)=(2\pi m!)^{-1/2}2^{-m/2}
({R\over\sqrt{2a_\al}})^m H_m(\mu_\al')\0
\b

Using intermediate basic functions we get for joint probability:
\a
P(\lm_\al,\lm_\beta)=
\int (\prod_{i=2}^N d\lm_i^{(\al)}d\lm_i^{(\beta)})
(\prod_{i=1}^N  \prod_{\gamma=\al+1}^{\beta-1} d\lm_i^{(\gamma)})\times\0\\
\times\det_{ij}[\xi_i^{(\al)}(\lm_j^{(\al)})]
\det_{ij}[\eta_i^{(\beta)}(\lm_j^{(\beta)})]e^{S-S_\al-S_\beta'}
\b
Integrating over intermediate eigenvalues $d\lm_i^{(\gamma)},
\gamma=\al+1,\ldots \beta-1$ we obtain the joint probability of
two-matrix model for which we already know the result.Hence we
get the result (\ref{qprob}).
All derivation above is valid also for more general potentials ,
polynomial-like $V_\al(\tau)=\sum_{k=1}^{p_\al}t_k\tau^k$ or not.
The sufficient incredients are the coefficients of the $Q$-matrices.

\section{Star-matrix model}
\setcounter{equation}{0}
\setcounter{subsection}{0}

We study the star-matrix model with partition function:
\a
Z=\int \prod_{i=1}^N (d\lm_i^{(0)} \prod_{\al=1}^q d\lm_i^{(\al)})
\prod_{i<j}[(\lm_i^{(0)}-\lm_j^{(0)})
\prod_{\al=1}^q(\lm_i^{(\al)}-\lm_j^{(\al)})]\times\0\\
\times\exp(\sum_{i=1}^N (
V_0(\lm_i^{(0)}+\sum_{\al=1}^qV_\al(\lm_i^{(\al)})+\sum_{\al=1}^{q}
c_\al\lm_i^{(0)}\lm_i^{(\al)})
\b

We define the orthogonal polynomial basis as $\xi_n$ and (instead of one
conjugate
polynomial $\eta_m$  $q+1$ polynomials $\eta_m^{(\al)}$:
\a
\int d\lm^{(0)}\prod_{\al=1}^q d\lm^{(\al)}
\xi_n^q(\lm^{(0)})e^{V_0+\sum_{\al=1}^{q}(V_\al+
c_\al\lm^{(0)}\lm^{(\al)})}\prod_{\al=1}^q \eta_{m_\al}^{(\al)}(\lm^{(\al)})
=h_n\delta_{nm},\0\\ m=m_\al,\al=1,\ldots q.
\b
This basis is unusual but it works quite well at least for gaussian potentials:
$V_\al(\tau)=t_\al\tau^2+u_\al\tau,\al=0,1,\ldots q$

We introduce $Q$-matrices as:
\a
\int d\lm^{(0)}\prod_{\al=1}^q d\lm^{(\al)} \xi_n^q(\lm^{(0)})\lm^{(\al)}
 e^{V_0+\sum_{\al=1}^{q}(V_\al+
c_\al\lm^{(0)}\lm^{(\al)})}\prod_{\al=1}^q \eta_{m_\al}^{(\al)}(\lm^{(\al)})
=h_nQ_{\al,nm},\0\\ m=m_\al,\al=1,\ldots q.\0\\
\b

The coupling conditions are:
\a
q P_0+2t_0Q_0+u_0+\sum_{\al=1}^{q}c_\al Q_\al &=&0\0\\
{\overline P_\al}+2t_\al Q_\al+u_\al+c_\al Q_0 &=&0,\al=1,\ldots q
\b

With the following parametrization of $Q$-matrices:
\a
Q_0&=&I_++a_0I_0+a_1\epsilon_-\\
Q_\al&=&b_\al/R_\al I_++d_\al I_0+R_\al\epsilon_-,\al=1,\ldots q\0
\b
we arrive at following equations:
\a
2t_\al R_\al+c_\al a_1=0\0\\
2t_\al b_\al+n+c_\al R_\al=0\0\\
2t_\al d_\al+u_\al+c_\al a_0=0\0\\
2t_0+\sum{c_\al b_\al\over R_\al}=0\\
2t_0a_0+u_0+\sum c_\al d_\al=0\0\\
2t_0a_1+qn+\sum c_\al R_\al=0\0
\b
Solving the coupling conditions we get :
\a
a_1&=&-{2q\over A},a_0={1\over A}(\sum{c_\al u_\al\over t_\al}-2u_0)\0\\
b_\al&=&-{1\over 2t_\al^2}({c_\al^2 q\over A}+t_\al),
R_\al={c_\al q\over t_\al A}\0\\
d_\al&=&{1\over At_\al}(c_\al u_0-2t_0u_\al+u_\al\sum{c_\al^2\over 2t_\al}-
c_\al\sum{c_\al u_\al\over 2t_\al})\0
\b
where $A=4t_0-\sum c_\al^2 / t_\al$.

In the same way we get the basic functions for $Q$-matrix model we can obtain
them for
star matrix model:
\a
\xi_n(\lm^{(0)})&=& H_n(\lm'^{(0)}),\\
\eta_m^{(\al)}(\lm^{(\al)})&=&R_\al^n H_n(\lm'^{(\al)}),\al=0,1\ldots q\0
\b
with:
\a
\lm'^{(0)}={\lm^{(0)}-a_0\over \sqrt{2a_1}},
\lm'^{(\al)}={\lm^{(\al)}-d_\al\over \sqrt{2b_\al}}
\label{starcoef}
\b

Because these basic functions satisfy relation:
\a
\eta_m(\lm^{(0)})=\int e^{V_\al+c_\al \lm^{(0)}\lm^{(\al)}}\eta_m(\lm^{(\al)})
\b
we can integrate over Vandermonde determinants:
\a
\det_{ij}[\eta_i^{(0)}(\lm_j^{(0)})]=\int e^{V_\al+c_\al \lm^{(0)}\lm^{(\al)}}
\det_{ij}[\eta_i^{(\al)}(\lm_j^{(\al)})]
\b
Then we have for the joint probability of two eigenvalues the simple
expression:
\a
P(\lm^{(\al)},\lm^{(\beta)})\sim P_{Herm}(\lm'^{(\al)},\lm'^{(\beta)}),\
\al,\beta=0,1\ldots q
\b
with $\lm',\mu'$ given by equation (\ref{starcoef}).

\section{Generalized Calogero-Sutherland model}
\setcounter{equation}{0}
\setcounter{subsection}{0}

The connection with Calogero model permits the calculation of the joint
distribution functions for random multimatrix models for other ensembles,
different from the hermitean one.

We obtain the Calogero model related to the 2-matrix model .The eigenvalue
problem for
Calogero model follows from the heat equation satisfied by the Itzykson- Zuber
integral.

We introduce the kernel:
\a
K(X,Y| t)= <X|e^{-t D}|Y>=(2\pi t)^{-N^2/2}\int dU \exp[-{1\over 2t}Tr
(X-UYU^+)]
\label{kernel}
\b
which is related with the Itzykson-Zuber  integral $K(X,Y| t=1)=\exp(-{1\over
2t}Tr(X^2+Y^2))I(X,Y)$:
\a
I(X,Y)=\int dU \exp[Tr (XUYU^+)]={det_{ij}(e^{x_i y_j})\over
(\Delta(X)\Delta(Y))^{\beta/2}}
\b
The kernel (\ref{kernel}) satisfies the heat equation \cite{Mehta2}\cite{IZ}:
\a
({\d \over \d t}+D_X)\tilde{K}(X,Y|t)=\delta(X,Y)
\label{heat}
\b
where $\tilde{K}(X,Y|t)=(\Delta(X)\Delta(Y))^{\beta/2} K(X,Y| t)$ and the
laplacian is:
\a
D_X=-{1\over 2}\sum_i{\d^2 \over \d x_i^2}+{\beta\over 2}({\beta\over
2}-1)\sum_{i<j}{1\over (x_i-x_j)^2}
\b
Solving equation (\ref{heat}) gives:
\a
\tilde{K}(X,Y|t)=(2\pi t)^{-N^2/2}\sum_\sigma \eta_\sigma
\exp[-{1\over 2t}\sum_i(x_{\sigma(i)}-y_i)^2]
\b
from which follows the expression for the Itzykson-Zuber  integral  ($\sigma $
is the permutation).

We introduce the function:
\a
\Phi(X|t)=\int \tilde{K}(X,Y|t)\Phi(Y) dY
\b
that fulfills the heat equation with initial condition $\Phi(X|t=0)=\Phi(X) $.

We can search for stationary solutions in the form $\Phi(X|t)=\sum_n
\Phi_n(X)e^{-E_n t}$
where  $\Phi_n(X)$ satisfies the Calogero equation  (without potential term):
\a
\left(-{1\over 2}\sum_i{\d^2 \over \d x_i^2}+{\beta\over 2}({\beta\over
2}-1)\sum_{i<j}{1\over (x_i-x_j)^2}\right)\Phi_n(X)=E_n \Phi_n(X)
\b
The eigenvalues of matrix $X$ are chosen such that $y_1<y_2\ldots <y_N$. These
eigenvalues
$y_1\ldots y_N$ are mapped by the kernel $\tilde{K}(X,Y|t)$   into
$x_{\sigma(1)}\ldots x_{\sigma(N)}$.

For $t\rightarrow 0$, the kernel $\tilde{K}(X,Y|t)$  tends to $\sum_\sigma
\eta_\sigma \delta^{(N)}(x_{\sigma(i)}-y_i)$. Hence if we consider  $\Psi(X)$
as a particular solution of Calogero model with $x_1<x_2\ldots <x_N$ ,the
function $\Phi(X|t=0)$ is the general solution for eigenvalues $x_i$ in
arbitrary order, being the linear combination of functions
$\Psi(\sigma X)$ :
\a
\Phi(X|t=0)=\sum_\sigma \Psi(\sigma X), \  \Psi(\sigma X)=\eta_\sigma \Psi(X)\0
\b
where $\sigma $ is the permutation of eigenvalues $x_i$ ; $\eta_\sigma =-1$ for
free fermions
($\beta=2$ for hermitean matrices) and  $\eta_\sigma =+1$ for free bosons
($\beta \rightarrow 0$ for harmonic oscillator) .

For $t\rightarrow \infty$ the dominant contribution is given by the vacuum
configuration $\Phi_0(X)$.
The kernel $\tilde{K}(X,Y|t)$  plays the role of instanton propagator
connecting the initial vacuum
configuration $\Psi_0(Y)=(\Delta(Y))^{\beta/2}$ to final vacuum configuration
$\Phi_0(X)=(\Delta(X))^{\beta/2}$.

For 2-matrix model we can define the Generalized Calogero system:
\a
\left(-{1\over 2}(\sum_i{\d^2 \over \d \lm_i^2}+
\sum_i{\d^2 \over \d \mu_i^2})+
{\beta\over 2}({\beta\over 2}-1)\sum_{i<j}({1\over (\lm_i-\lm_j)^2}+
{1\over (\mu_i-\mu_j)^2})+\right.\0\\
\left. +\sum_i(V_1(\lm_i)+V_2(\mu_i)+c\lm_i \mu_i)\right)
\Phi_n(\lm)\Psi_m(\mu)=E_{nm} \Phi_n(\lm)\Psi_m(\mu)
\label{gencal}
\b

When $c=0$ the generalized system splits into two Calogero systems :
\a
\left(-{1\over 2}\sum_i{\d^2 \over \d \lm_i^2}+{\beta\over 2}({\beta\over
2}-1)\sum_{i<j}{1\over (\lm_i-\lm_j)^2}+\sum\lm'^2\right)\Phi_n(\lm)=E_{1,n}
\Phi_n(\lm)\\
\left(-{1\over 2}\sum_i{\d^2 \over \d \mu_i^2}+{\beta\over 2}({\beta\over
2}-1)\sum_{i<j}{1\over (\mu_i-\mu_j)^2}+\sum\mu'^2\right)\Phi_n(\mu)=E_{2,m}
\Phi_n(\mu)\0
\label{2cal}
\b
The ground states can be written in terms of the eigenfunctions (\ref{pol}):
\a
\Phi_0(\lm)=(det_{ij }\xi_i(\lm_j))^{\beta/2}\exp(-\sum_i \lm_i'^2/2)\0\\
\Psi_0(\mu)=(det_{ij }\eta_i(\mu_j))^{\beta/2}\exp(-\sum_i \mu_i'^2/2)
\label{ampl}
\b
We can see that the probability of amplitudes (\ref{ampl}) is the partition
function of the 2-matrix model:
\a
Z=\int d\lm_1 \ldots d\lm_N |\Phi_0(\lm)|^2=\int d\mu_1 \ldots d\mu_N
|\Psi_0(\mu)|^2\0
\b

The system (\ref{2cal}) permits us to calculate the joint probability
$P(\lm,\mu)$ for general
ensemble. It coincides with formula (\ref{2density'}) (for $c=0$) where we
replace $N$ by $\beta N/2$:
\a
P(\lm,\mu)=\sqrt{\beta N/2-\lm'^2}\sqrt{\beta N/2-\mu'^2}
\b

\section{Conclusions}

These models present interest in the study of quantum chaos for $q$ systems
interacting in various ways.
The density of levels depends on the total energy which behaves like $N$, for
large $N$.
The interaction of $q$ subsystems redistribute  the energy between the
subsystems and change in non-trivial way the joint distribution functions.
Different kinds of interaction (chain or star- type ) give different
probabilities for energy levels.

\begin{center}
AKNOWLEDGEMENTS
\end{center}
I  would  like  to  thank  Prof.L.Bonora  and Dr.F.Nesti  for  many  usefull
discussions. I also want to aknowledge the SISSA  staff  and  students
for the stimulating enviroment.

\begin{center}
FIGURE CAPTIONS
\end{center}
\noindent Figure 1. -Represents the level density $P(x,x)$ in terms of the
energy $x=\al\lm$ and the asymmetry $y=N\epsilon$.For $y=0$
we have the semicircular law $P(x,x)=\sqrt{2N-x^2}$ and for small $y \neq 0$ we
get the oscilations of level density.
\end{document}